\input psfig

\documentstyle[twoside,fleqn,espcrc2]{article}

\newcommand{\AmS}{{\protect\the\textfont2
  A\kern-.1667em\lower.5ex\hbox{M}\kern-.125emS}}

\title{Physical Effects of Infrared Quark Eigenmodes in LQCD
  \thanks{Talk presented by E.~Eichten}}
\author{ A.~Duncan\address{Dept. of Physics and Astronomy,University of Pittsburgh,
 Pittsburgh, PA 15260},%
  E.~Eichten\address{ Fermilab, PO Box 500, Batavia, IL60510},%
  and 
  H.~Thacker\address{Dept. of Physics, University of Virginia, 
 Charlottesville, VA 22901}}
\begin{document}
\begin{abstract}
A truncated determinant algorithm is used to study the physical
effects of the quark eigenmodes associated with eigenvalues below 400 MeV. 
This initial study focuses on coarse lattices 
(with O($a^2$) improved gauge action), 
light internal quark masses and large physical volumes.
Four bellwether full QCD processes are discussed: 
topological charge distributions, the eta prime propagator,
string breaking as observed in the static energy and
the rho decay into two pions.
\end{abstract}

\maketitle

\section{Basics}

In the truncated determinant approach (TDA) to 
full QCD,  the quark determinant, $ {\cal D}(A) = det(H) =
det(\gamma_{5}(D\!\!\!/(A)-m))$ is split-up gauge invariantly 
into an infrared part and an ultraviolet part\cite{truncdet}.
\begin{equation}
\label{eq:split}
{\cal D}(A)={\cal D}_{IR}(A){\cal D}_{UV}(A)  
\end{equation}
The ultraviolet part, ${\cal D}_{UV}$, can be accurately fit 
by a linear combination of a small number of Wilson loops\cite{looppaper}.
The infrared part ${\cal D}_{IR}(A)$ is defined as the product of 
the lowest $N_{\lambda}$ positive and negative eigenvalues of $H$, 
with $|\lambda_{i}|\leq \Lambda_{QCD}$ (typically, $\simeq$ 400-500 MeV). 
The eigenvalues $\lambda_{a}$ of $H$ are gauge invariant 
and measure quark off-shellness.
The cutoff (for the separation in Eq. \ref{eq:split}) is tuned to include as much as
possible of the important low-energy chiral physics 
of the unquenched theory while leaving the fluctuations 
of $\ln {\cal D}_{IR}$ of order unity after each sweep updating
all links with the pure gauge action (assuring a tolerable acceptance
rate for the accept/reject stage)\cite{duncan99}. 
This procedure works well even for kappa values 
arbitrarily close to kappa critical. 

Initial studies using TDA focus on the qualitative 
physical effects of the inclusion of the infrared quark eigenmodes.
For this purpose, coarse lattices with large physical volumes 
are appropriate. 

\section{Bellwethers}

\begin{table}
\caption{\noindent Coarse lattices studied.}
\label{tbl:cases}
\begin{center}
\begin{tabular}{|c|c|}
\hline
Label & Volume; Fermions; Gauge Action \\
PW6  & $6^4$; $n_{cut} = 840$ $\kappa = .2180$; \\
     & Naive $\beta = 4.5$ \\ 
RW6  & $6^4$; $n_{cut} = 840$ $\kappa = .1950$; \\
     & $O(a^2)$ $\beta = 6.8$\\
RW8  & $8^4$; $n_{cut} = 1780$ $\kappa = .1950$; \\
     & $O(a^2)$ $\beta = 6.8$\\
\hline
\end{tabular}
\end{center}
\end{table}

The coarse lattices given in Table \ref{tbl:cases}
are being studied on PC clusters.
The $O(a^2)$ improved gauge action, $\beta = 6.8[1.0(plaq) 
- 0.08268(rect) - 0.01240(para)]$, was adjusted in Ref. \cite{impgauge}
to have approximately the same lattice scale $a = 0.4 fm$ as the 
naive gauge action at $\beta = 4.5$.  The physical lattice size is
2.4 (fm) for PW6 and RW6 and 3.2 (fm) for RW8; $\kappa_{c} = .2190$ for 
PW6, and $.1960$ for RW6 and RW8; and
eigenvalue cutoff scale is 560 MeV for PW6 and RW6, and 445 MeV for RW8.
In addition, $10^3\times 20$ lattices at $\beta = 5.7$, $c_{SW} = 1.57$ and 
$n_{cut} = 520$ ($E_{cut} = 460 MeV$) are being generated on ACPMAPS.

\begin{figure}
\psfig{figure=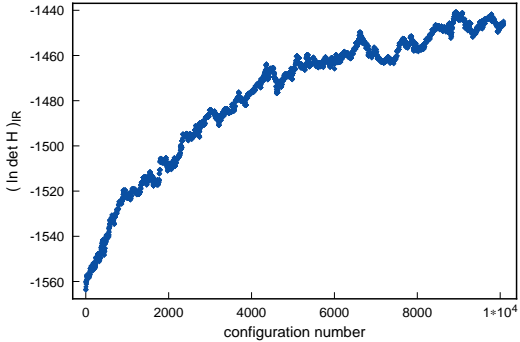,
width=0.95\hsize}
\vspace{-0.35in}
\caption{Reaching equilibrium for the PW6 lattices.}
\label{fig:detj}
\end{figure}

As shown in Figure \ref{fig:detj} it took about 10,000 full steps 
for the PW6 lattice configurations to equilibrate (reflecting critical
slowing down - a few hundred suffice on small volumes). 

Four bellwethers can be used to characterize the physical differences 
between quenched and full QCD. They are discussed (in order of 
increasing difficulty to
observe in lattice calculations) in the following four subsections.

\subsection{Topolopy}
The topological charge, $Q_{TOP}$, can be expressed 
in terms of the eigenvalues of the Wilson-Dirac operator. 
\begin{equation}
 Q_{TOP} = {1 \over 2\kappa} (1 - {\kappa \over\kappa_{c}}) \sum_{i=1}^{N} {1 \over \lambda_i}
\end{equation}
This sum is quickly saturated by the low eigenvalues.

In full QCD configurations with very small eigenvalues of $H$
are suppressed by the quark determinant factor. 
In particular, non-zero topological charges must be suppressed
in the chiral limit ($m_q \rightarrow 0$). 
Furthermore,  the functional dependence of the topological charge distribution, $P_Q$,  
on the light quark mass $m_q$ is predicted by the chiral analysis of
Leutwyler and Smilga\cite{LeutSmil}.  
\begin{equation}
 P_Q = I_Q(x)^2 - I_{Q+1}(x) I_{Q-1}(x) \label{eq:LS} 
\end{equation}
where 
\begin{equation}
 x = 1/2 V f^2_{\pi}m^2_{\pi} = V m_q <\overline \psi \psi >,
\end{equation}
$I_{Q}$ are modified Bessel Functions of order Q and V is the total space-time volume.

The quantitative agreement with the expected behaviour of Eq. \ref{eq:LS} has already
been reported for the TDA method\cite{truncdet}.  General agreement is also 
observed on the coarse lattices of the present studies. 

\subsection{Eta Prime Mass}

\begin{figure}
\psfig{figure=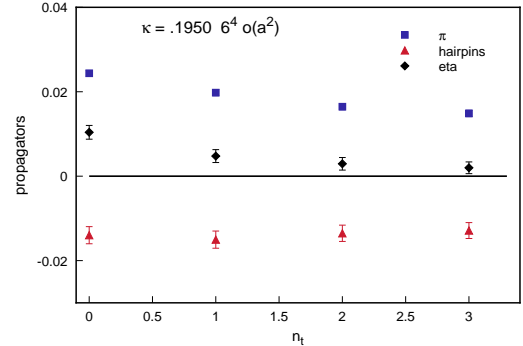,
width=0.95\hsize}
\vspace{-0.35in}
\caption{Eta prime propagator for the RW6 lattices. 
The valence term (squares), hairpins term (triangles)
and total propagator (diamonds) are shown separately.}
\label{fig:etap}
\end{figure}

The relation between the axial $U(1)$ anomaly and the $\eta '$ mass 
is well understood in full QCD. 
For two light quarks ($N_f = 2$), 
$m^2_{\eta} = m^2_{\pi} + m^2_0$ where 
$m^2_0 = 2N_f\chi/f^2_{\pi}$ and the topological susceptibility is  
$V\chi \equiv <Q_{TOP}^2>_{\rm quenched}$.
The full $\eta$ propagator is the sum of a connected (valence quark) term and a
disconnected (hairpin) term.  Thus, in the continuum, the momentum space 
full propagator can be written: 
\begin{eqnarray}
 \lefteqn{(p^2 + m^2_{\pi} + m^2_0)^{-1}  =  ~~(p^2 + m^2_{\pi})^{-1}~-~ } 
               \\
  & & ~~~~~~~m^2_0(p^2 + m^2_{\pi})^{-1} 
             (p^2 + m^2_{\pi} + m^2_0)^{-1}   \nonumber 
\end{eqnarray}
These separate terms and their sum are shown in Figure \ref{fig:etap} for the
RW6 lattices. 
The cancellation between the valence and hairpin terms in the full propagator is evident.

\subsection{Static Energy}
 
To date no convincing evidence for string breaking in full QCD has been presented using
calculations of the static energy alone. However string breaking has been seen using
the TDA method in 2D QED\cite{truncdet} and by the HMC method in 3D QCD\cite{3dQCD}.
Studies of the $\overline cc$ and $\overline bb$ systens, lead to the expectation
that virtual pair effects (below heavy-light meson pair production threshold) will
soften the linear rise in the static energy, while above threshold the potential
will flatten out (i.e.) the string will break.
The heavy-light 
meson mass is $0.81 \pm .02$ for the RW6 lattices.   

\begin{figure}
\psfig{figure=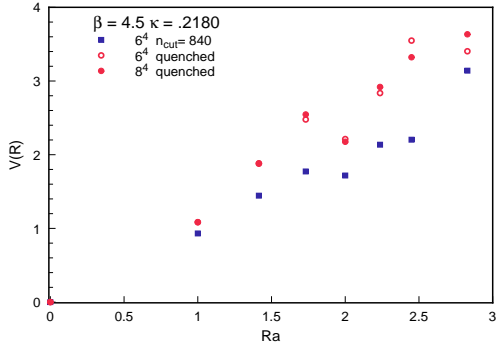,
width=0.95\hsize}
\vspace{-0.35in}
\caption{Comparison of the static energy between the RW6 lattices 
(solid squares) and unquenched $6^4$ (open circles) (and 
$8^4$ (solid circles)) lattices at $\beta = 4.5$.  
The static energy was extracted from the ratio of time 
slices $T=2/T=1$.} 
\label{fig:wilson}
\end{figure}

Figure \ref{fig:wilson} shows the static energy for 200 RW6 lattices versus 
the same number of unquenched $6^4$ lattices at $\beta = 4.5$. 
There is evidence of screening from the virtual pairs but no hard evidence
of string breaking is found. Seeing string breaking 
will require more statistics (to study
$(T=3)/(T=2)$ with small error bars) and also the study of the $RW8$ lattices. 

\subsection{Vector Meson Resonances}

\begin{figure}
\psfig{figure=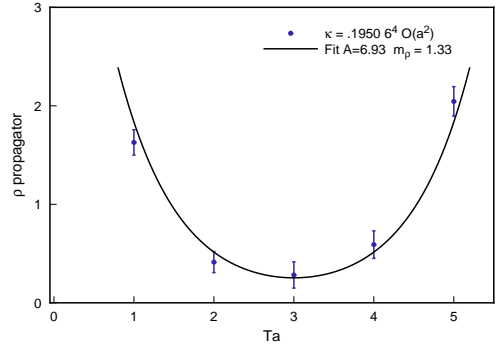,
width=0.95\hsize}
\vspace{-0.35in}
\caption{Rho propagator for the RW6 lattices.}
\label{fig:rho}
\end{figure}

For the RW6 and RW8 lattices at $\kappa = .1950$,
the rho mass is 1.33 and the pion mass is 0.205 (in lattice units);
hence the $\pi/\rho$ mass ratio is close to the physical value.
For example, the rho propagator for the RW6 lattice 
is shown in Figure \ref{fig:rho}.
However, since this is a P wave coupling, the physical volume of the
lattice must be large enough that the decay is allowed with the
first nonzero momentum, $p_{min} = {2 \pi \over Na}$. This requires
a $10^4$ lattice (RW10) or creating a rho with initial momentum $p_{min}$.
Neither of these alternatives have been studied as yet.

\section{Status}

The present status of full QCD bellwethers is as follows:
\begin{itemize}
\item The behaviour of the topological charge distribution $Q^2$ 
as a function of light quark mass $m_q$\cite{LeutSmil} -- Seen.  
\item The eta prime mass - $m_{\eta '}^2 = m_{\pi}^2 + m_0^2$ -- Seen.
\item The static energy - string breaking. -- In progress but needs more statistics.
\item Vector meson resonances - $\rho \rightarrow \pi \pi$. -- 
Yet to be studied in detail. 
\end{itemize}
Results for these four bellwether processes on coarse 
lattices should be available within a few months.

\vspace{-0.1in}


\end{document}